\documentstyle[12pt,epsf]{article}
\newcommand{\mathbold}[1]{\mbox{\rm\bf #1}}

\newcommand{\N}{{\cal N}}

\newcommand{\beq}{\begin{equation}}
\newcommand{\eeq}{\end{equation}}
\newcommand{\bea}{\begin{eqnarray}}
\newcommand{\eea}{\end{eqnarray}}

\newcommand{\gtrsim}{\ \rlap{\raise 2pt\hbox{$>$}}{\lower 2pt \hbox{$\sim$}}\ }
\newcommand{\lessim}{\ \rlap{\raise 2pt\hbox{$<$}}{\lower 2pt \hbox{$\sim$}}\ }

\newcommand{\ea}{{ et al.}}

\newcommand{\np}[1]{Nucl. Phys. {\bf #1}}
\newcommand{\pl}[1]{Phys. Lett. {\bf #1}}
\newcommand{\pr}[1]{Phys. Rev. {\bf #1}}
\newcommand{\prl}[1]{Phys. Rev. Lett. {\bf #1}}
\newcommand{\zp}[1]{Z. Phys. {\bf #1}}
\newcommand{\prep}[1]{Phys. Rep. {\bf #1}}

\newcommand{\mpl}[1]{Mod. Phys. Lett. {\bf #1}}
\newcommand{\ptp}[1]{Prog. Theor. Phys. {\bf #1}}
\newcommand{\arns}[1]{Ann. Rev. Nucl. Sci. {\bf #1}}
\newcommand{\anfis}[1]{An. F\'\i s. {\bf #1}}
\makeatletter
\if@twoside
\else \oddsidemargin 00pt \evensidemargin 00pt
\fi
\let\@eqnsel = \hfil

\expandafter\ifx\csname mathrm\endcsname\relax\def\mathrm#1{{\rm #1}}\fi
\@ifundefined{mathrm}{\def\mathrm#1{{\rm #1}}}{\relax}

\makeatother

\begin{document}
\thispagestyle{empty}
\null
\hfill CERN-TH/95-33

\hfill FTUV/95-10,IFIC/95-10

\hfill Lund-MPh-95/03

\hfill hep-ph/9503228

\vskip 1.5cm

\begin{center}
{\Large \bf      
NON-DECOUPLING OF HEAVY NEUTRINOS \par
AND LEPTON FLAVOUR VIOLATION
\par} \vskip 2.em
{\large		
{\sc D. Tommasini,}  \\[1ex] 
{\it Theory Division, CERN, CH-1211, Geneva 23,
Switzerland} \\[2ex]
{\sc G. Barenboim, J. Bernab\'eu,}  \\[1ex] 
{\it Departament de F\'\i sica Te\`orica, Universitat
de Val\`encia, and I.F.I.C.}\\
{\it 46100 Burjassot, Valencia, Spain} \\[1ex]
and \\[1ex]
{\sc C. Jarlskog}  \\[1ex] 
{\it Dept. of Math. Phys., LTH, Univ. of Lund, S-22100 Lund, Sweden} \\
\vskip 0.5em
\par}
\end{center} \par
\vfil
{\bf Abstract} \par
We consider a class of models predicting new heavy
neutral fermionic states, whose mixing with the light neutrinos can
be naturally significant and produce observable effects below the threshold
for their production. We update
the indirect limits on the flavour non-diagonal mixing parameters that can
be derived from unitarity, and show that significant rates are in general
expected for one-loop-induced rare processes due to the exchange of virtual
heavy neutrinos, involving the violation of the muon and electron lepton
numbers.
In particular, the amplitudes for $\mu$--$e$  conversion in nuclei and
for $\mu\to ee^+e^-$ show a
non-decoupling quadratic dependence on the heavy neutrino mass $M$, while
$\mu\to e\gamma$ is almost independent of the heavy scale above the
electroweak scale. These three processes are then used to set
stringent constraints on the flavour-violating mixing angles.
In all the cases considered, we point out explicitly that the
non-decoupling behaviour is strictly related to the spontaneous breaking
of the SU(2) symmetry.

\par
\vskip 0.5cm
\noindent CERN-TH/95-33 \par
\vskip .15mm
\noindent 3 March 1995 \par
\null
\setcounter{page}{0}
\clearpage

\phantom{\cite{fermionmix}\cite{fitnu}\cite{mix94}}
\phantom{\cite{see-saw}}
\phantom{\cite{lim-inami}\cite{cheng-li}}
\phantom{\cite{numix}\cite{mohapatra-valle}}
\phantom{\cite{ll2}\cite{unconventional}}

\section{Introduction}

New heavy neutral fermions could affect low-energy measurements,
made well below their production thresholds.
In general, these new states mix with the three light neutrinos,
thus modifying their neutral current (NC) and charged current (CC)
couplings.
Such effects depend on the light--heavy neutrino mixing angles,
which can then be constrained by the set of NC and CC precision data at
``low" energy \cite{fermionmix}--\cite{mix94}.
If the explanation for the smallness of the known neutrino masses
is given by the see-saw mechanism \cite{see-saw}, these mixing angles are
proportional to the square roots of the ratios of the light and the heavy
mass eigenvalues, so that their effects on the light neutrino couplings to
the standard gauge bosons vanish when the heavy neutrino masses go to
infinity. In this limit, the heavy neutrinos completely decouple from the
low-energy physics. In fact, not only the tree-level light--heavy mixing
effects, but also the loop diagrams involving
the heavy neutrinos, contributing to physical observables,
turn out to be suppressed by inverse powers of the
heavy mass scale \cite{lim-inami,cheng-li}.

However, the see-saw mechanism is not the only possible scenario to explain
the lightness of the known neutrinos. In particular, a
viable alternative involving heavy neutral states has been
considered \cite{numix}--\cite{unconventional},
where vanishingly small masses for the known
neutrinos are predicted by a symmetry argument,
and at the same time large light--heavy mixing angles are allowed.
In this case, due to the mixing effects in the light neutrino interactions,
the new neutral fermions affect the low-energy physics even
if their masses are very large.

In the present paper,
we discuss the conditions under which the mixing angles
between the standard and new neutral fermions heavier than $M_Z$
can be large, keeping at the same time the masses
of the known neutrinos below the laboratory limits.
We will then study the limits that can be set on the mixing parameters,
concentrating on the ones which would induce Lepton Flavour Violation
(LFV).
In particular, we will consider a class of models that allow total
lepton number conservation, and we will show that loop contributions
involving virtual heavy neutrinos
to the decays $\mu\to e\gamma$, $\mu\to ee^+e^-$, and to the
process of $\mu$--$e$  conversion in nuclei,
can be large and then provide significant constraints on the mixing
parameters.
In particular for a heavy mass scale around the electroweak scale
the induced $\mu\to e e^+ e^-$ and $\mu$--$e$  conversion rates
{\it increase} with the masses of the heavy neutrinos,
showing a non-decoupling behaviour,
and the present limits on the processes put significant
constraints in the space of the
light--heavy mixing parameters and new neutrino masses.
The planned experiments looking for muon--electron conversion
are specially suitable for finding signals arising from this kind of physics.

The paper is organized as follows.
In section 2, we discuss the class of models for neutrino mass
that predict significant light-heavy
mixing. In section 3, we review the direct
constraints on the flavour-diagonal, and up-date
the resulting indirect limits on the flavour-changing, mixing parameters.
In section 4, we compute the non-decoupling contributions
to the decays $\mu\to e\gamma$, $\mu\to ee^+e^-$, and to the
process of $\mu$--$e$  conversion in nuclei. We discuss
their importance and their interplay in constraining the models, and
we show how the non-decoupling behaviour is strictly related to the
breaking of the electroweak symmetry.
Finally, section 5 summarizes our conclusions.

\section{Models for light--heavy neutrino mixing}

Let us first
consider a generic extension of the Standard Model (SM), including
right-handed neutrinos (singlets under the SM group) as
the only new neutral fermions.
We can arrange all the independent neutrino degrees of
freedom in two vectors of {\it left-handed} fields,
$\nu$ and $N\equiv C\bar\nu_R^T$, where $C$ is the charge-conjugation
matrix and a family index is understood.
In the basis $(\nu,N)$ the mass matrix can be written in a block form as
\beq
{\mathbold M}=\pmatrix{m_{\nu \nu} & m_{\nu N}\cr
             m_{\nu N}^T & M_{N N}\cr}.
\label{nondec1}
\eeq

The entry $m_{\nu\nu}$ can be due to a possible lepton-number-violating vacuum
expectation value (VEV) of a triplet Higgs field, as in left--right models.
Although in principle the singlet neutrinos can be light, we are interested
in the case when all the new (i.e. non-SM) states are heavier than $\sim M_Z$.
This is also the theoretical expectation in most models, which generally
predict large masses for the heavy states. In the
one family case, as far as the entry $m_{\nu\nu}$ can be neglected,
we have the usual see-saw mechanism \cite{see-saw} for the
generation of a small neutrino mass.  In this case, the
light--heavy mixing angle $\theta$
depends on the ratio of the light and heavy mass scales,
as $\sin^2\theta \sim m/M$ ($m\sim m_{\nu N}^2/M_{N N}$,
$M\sim M_{N N}$).
For $M\gtrsim M_Z$, taking the laboratory limits on the $\nu_e$,
$\nu_\mu$ and $\nu_\tau$ masses \cite{pdg94}, we get the bounds
$\sin^2\theta_{\nu_e}\lessim 10^{-10}$,
$\sin^2\theta_{\nu_\mu}\lessim 10^{-6}$,
$\sin^2\theta_{\nu_\tau}\lessim 10^{-3}$, which are too small to have
any phenomenological interest.
To have significant light--heavy mixing in the one family
case, we are left with only one solution, requiring
that the entry $m_{\nu\nu}$ be non vanishing, and satisfying the
relation $m_{\nu\nu}=m_{\nu N}^2/M_{N N}$. This would
ensure that the mass matrix of Eq. (\ref{nondec1}) be singular,
so that the mixing angle $\sin\theta\sim m_{\nu N}/M_{N N}$
would no longer be related to the ratio of the light to the heavy eigenvalues,
and would be allowed to be as large as $O(1)$.
However, it seems hard to find a reasonable motivation for the
underlying fine tuning of the parameters in the mass matrix.
The model with three families of left- and right- handed neutrinos
($\nu\equiv(\nu_e,\nu_\mu,\nu_\tau)$, $N\equiv (N_e,N_\mu,N_\tau)$)
allows a solution even in the case $m_{\nu\nu}=0$ \cite{buchmuller-greub}.
In this case, the fine-tuning conditions, which allow for finite
light--heavy mixing and vanishing mass of the known neutrinos, are
the following: 1) the Dirac mass
matrix $m_{\nu N}$ is of rank 1, that is all the three lines
(rows) are proportional; 2) the trace $Tr(m_{\nu N}^T M_{NN}^{-
1} m_{\nu N})=0$ (assuming that $M_{NN}$ is not singular).

These considerations can be generalized. The neutrino mass matrix should
be singular with a three times degenerate zero eigenvalue,
in the limit in which the masses of the three known
neutrinos are neglected. In the see-saw mechanism, this is ensured by
letting the heavy scale go to infinity, which implies that the light--heavy
mixing angles go to zero. However, if for some reason the mass matrix is
(three times) singular even for {\it finite} values of the heavy scale,
the light--heavy mixing angle can be substantial. Any model realizing this idea
in a natural way, e.g. due to a symmetry argument,
is a viable alternative to the see-saw mechanism to
explain the lightness of the known neutrinos.

For instance, let us assume
that pairs $N$, $N'$ of (left-handed) new neutrinos exist,
with the lepton-number assignments $L(N)=-L(N')=L(\nu)=1$,
and that $L$ is conserved. We understand a family index, i.e.
$\nu\equiv(\nu_e,\nu_\mu,\nu_\tau)$, $N\equiv (N_1,...,N_{n-3}),$
$N'\equiv(N'_1,...,N'_{n-3}),$ where $n-3$
is the number of new pairs of neutral
fermions. Then, in the basis $(\nu, N, N')$, the mass matrix is
\beq
{\mathbold M}=\pmatrix{0 & 0 & M_{\nu N'}\cr
             0 & 0 & M_{N N'}\cr
M_{\nu N'}^T & M_{NN'}^T & 0\cr},
\label{nondec2}
\eeq
which is singular and ensures that three eigenstates form
massless Weyl neutrinos. In fact, as in the SM, the light states remain
with no chirality partners and hence massless.\footnote{Small
$L$-violating Majorana mass terms
for the states $\nu$ and $N$ could also be allowed \cite{ll2},
and could be relevant for explaining
the solar neutrino deficit via neutrino oscillations.}
On the contrary, the heavy states form Dirac neutrinos,
whose left-handed components
are mainly the $N$ and whose right-handed parts are given by
$C\bar (N')^T$. Mass matrices of the form (\ref{nondec2})
have been considered in Ref. \cite{numix}.
They can arise in generalized E$_6$ models
\cite{mohapatra-valle}--\cite{unconventional},
as well as in models predicting other kinds of
vector multiplets
(singlets, triplets, \dots) or new  mirror multiplets of
leptons \cite{maalampi-roos} with neutral components $N$, $N'$.

The mass matrix
(\ref{nondec2}) can be put in a ``Dirac diagonal" form by an ``orthogonal"
transformation,
\beq
{\mathbold U}^T {\mathbold M} {\mathbold U}=\pmatrix{0 & 0 & 0\cr
             0 & 0 & M\cr
             0 & M & 0\cr},
\label{nondec3}
\eeq
where the block $M$ is diagonal.
The {\it unitary} matrix ${\mathbold U}$ in Eq. (\ref{nondec3})
can be chosen in the form
\beq
{\mathbold U}=\pmatrix{A&G&0\cr
             F&H&0\cr
             0&0&K\cr}
.\label{nondec4}
\eeq
Several relations amongst the blocks in
(\ref{nondec4}) can also be deduced from the unitarity condition
${\mathbold U}^\dagger{\mathbold U}={\mathbold U}{\mathbold U}^\dagger=
{\mathbold 1}$.
Equation (\ref{nondec4}) describes the mixing between the $\nu$ and $N$ states,
the mixing parameters being the elements of the matrix
\beq
G H^{-1}=-(F A^{-1})^\dagger = [M_{\nu N'}M_{N'N'}^{-1}]^*
.\label{nondec4b}
\eeq
Clearly, if for the relevant matrix elements $M_{\nu N'}\sim M_{N N'}$, the
mixings between $\nu$ and $N$ can be arbitrarily large.
We will consider in the following two particular cases:

i) The new neutrinos $N$ are {\it ordinary}, i.e. they belong
to a weak (left-handed) doublet. Then
$M_{\nu N'}$ and $M_{N N'}$ could be generated by
vacuum expectation values of Higgs fields
transforming in the same way under SU(2)
so that the $\nu$--$N$ mixing could be naturally close to maximal.
In particular, we can consider the SM with $n>3$
families, and with $n-3$ right-handed neutrinos.
Then $N$ would describe the neutrinos of the new $n-3$
families, appearing with right-handed partners $\overline{N'}$.
In this picture, the three known neutrinos remain
strictly massless since they have no right-handed component.
The same scenario arises in some lepton-number-conserving
E$_6$ models \cite{unconventional}, predicting three new ordinary
$N$ states (one per standard family) and six isosinglets, three of which
can play the role of our $N'$ in such a way that the relevant part of the
mass matrix assumes the form of Eq. (\ref{nondec2}).

ii) Another case that has been considered in the literature
\cite{mohapatra-valle} corresponds to both $N$ and $N'$ singlets. In this
case, a significant light--heavy mixing can be expected only if the
SU(2)-invariant mass term $M_{NN'}$ is
generated not far from the electroweak scale.

In both the above cases, the $N'$ states are assumed to be isosinglets.
This implies that the block $M_{\nu N'}$ violates SU(2), and can be
generated by the VEV of a doublet Higgs field at the electroweak scale.

\section{Constraints from flavour-diagonal processes}

The mixing with heavy states would affect the observables at energies
below the threshold for their production.
Following e.g. Ref. \cite{fitnu}, we introduce a vector
${N}$ to describe all the new independent
neutral fermionic degrees of freedom which mix with the three known
left-handed neutrinos $\nu\equiv(\nu_e,\nu_\mu,\nu_\tau)$.
We will use only left-handed fields, without
distinguishing between neutrinos and antineutrinos.
Then the light $n$ and heavy $\N$ mass eigenstates
can be obtained by a unitary transformation
\beq
\pmatrix{\nu\cr
         {N}\cr}=
\pmatrix{A&G\cr
             F&H\cr}
\pmatrix{n\cr
         \N\cr}
.\label{nondec5}
\eeq
This general formalism covers in particular the cases discussed in the
previous section. In the case of Eqs. (\ref{nondec2})--(\ref{nondec4}), this
means that we are now focusing on the submatrix involving the light--heavy
mixing, given by Eq. (\ref{nondec5}), which is unitary.
The light--heavy mixing is described by the matrix $G$, and is reflected also
in the non-unitarity of the block $A$ ($AA^\dagger+GG^\dagger=1$).
Notice that $A$ also describes the leptonic
Cabibbo--Kobayashi--Maskawa mixing, in the basis where
the mass matrix for the (light) charged leptons is diagonal.

In processes occurring at energies below the threshold for the production of
the heavy states, the standard gauge
eigenstate $\nu_a$ ($a=e,\mu,\tau$) is effectively replaced by its
(normalized) projection $\vert \nu_a^{light}\rangle$
onto the subspace of the light neutrinos $\vert n_i\rangle$
($i=1,2,3$),
\beq
\vert \nu_a^{light}\rangle \equiv {1\over
c_{\nu_a}}\sum_{i=1}^{3} A^\dagger_{ia}
\vert n_i\rangle,
\label{nondec6}
\eeq
where $c_{\nu_a}^2\equiv\cos^2\theta_{\nu_a}\equiv (AA^\dagger)_{aa}$.
The state $\vert \nu_a^{light}\rangle$ has
non-trivial projections on the
subspace of the standard neutrinos $\vert \nu_b\rangle$
as well as on the subspaces
of the new neutrinos $\vert {N}_B\rangle$.
In fact we have
\beq
\begin{array}{ll}
{\displaystyle
\sum_b}\vert\langle\nu_b\vert\nu_a^{light}\rangle\vert^2=
&{(AA^\dagger)_{aa}^2\over c_{\nu_a}^2}
=c_{\nu_a}^2 ,\\
{\displaystyle
\sum_B}\vert\langle{N}_B\vert\nu_a^{light}\rangle\vert^2=
&{(AF^\dagger FA^\dagger)_{aa}\over c_{\nu_a}^2}=s_{\nu_a}^2,
\end{array} \label{nondec7}
\eeq
with $s_{\nu_a}^2\equiv 1-c_{\nu_a}^2=\sin^2\theta_{\nu_a}$.
The parameter $\theta_{\nu_a}$ measures the total amount of mixing
of the known state of flavour $a=e,\mu,\tau$ with the new states.
These three mixing angles are sufficient to describe the {\it tree-level}
effects of the light--heavy mixing in the CC and NC processes
at energies
below the threshold for the production of the heavy states \cite{fitnu}.

The entries of the matrix $GG^\dagger$, describing the mixing
with the new neutrinos, are limited by the constraints on
CC universality and, if the heavy states do not
belong to SU(2) doublets, by the measurement of the $Z$ boson invisible
width at LEP \cite{fermionmix}--\cite{mix94}.
For the diagonal elements $(GG^\dagger)_{aa}\equiv s^2_{\nu_a}$,
the 90\% C.L. bounds are \cite{fitnu,mix94}
\beq
s^2_{\nu_e}<0.007(0.005),\qquad s^2_{\nu_\mu}<0.002
,\qquad s^2_{\nu_\tau}<0.03(0.01),
\label{nondec8}
\eeq
where the more conservative limits are due to the CC constraints
and apply to any kind of heavy neutrinos, while the limits in
parentheses correspond to the mixing with SU(2) singlets and take into
account also the LEP \cite{lep94} and SLC \cite{alr-slc} data.
For a complete discussion, we refer to \cite{fitnu,mix94}.
These limits can be somewhat
relaxed if the cancellations with the effects due to different
fermion-mixing parameters which might be present in some extended models
are taken into account \cite{fermionmix}--\cite{mix94}.
Nevertheless, we will not try to allow for
the corresponding fine-tunings and we
will consider the stringent bounds of Eq. (\ref{nondec8}) to be reliable.
For the off-diagonal elements of
the matrix $GG^\dagger$, indirect limits
can be obtained from Eq. (\ref{nondec8}) and the relation \cite{ll2}
\beq
\vert (GG^\dagger)_{ab}\vert<\vert s_{\nu_a}s_{\nu_b}\vert,
\label{nondec9}
\eeq
which can be deduced from the unitarity of the full mixing matrix by
applying the Schwartz inequality.
Using the bounds of Eq. (\ref{nondec8}), we find that all the elements of the
matrix $GG^\dagger$ can be constrained as in Table 1.

Loop diagrams involving virtual heavy neutrinos
can contribute to flavour-diagonal observables
\cite{lim-inami,nuloopfcons}, such as the leptonic widths $Z\to l\bar l$
or the polarization asymmetries measured at the $Z$ peak \cite{lep94,alr-slc}.
However, taking into account the
updated limits of Table 1, the predictions for these observables turn out
to be below the attainable experimental limits.

Heavy neutrinos in general also affect the electroweak radiative
corrections which are tested e.g. in the LEP experiments. For instance, if
the new neutral states $N$ belong to SU(2) doublets $\pmatrix{N\cr E}$,
they will then contribute to the $\rho$ parameter \cite{rho-mt,rhoparam}
through loop diagrams, resulting in a (top-like) non-decoupling dependence,
\beq
\delta\rho\simeq\sum {G_F\over 8\sqrt2\pi^2}\Delta M^2,
\label{nondec10}
\eeq
where $\Delta M^2\equiv M_E^2+M_N^2 -
{4 M_E^2 M_N^2\over M_E^2 - M_N^2} \ln {M_E\over M_N}\ge
(M_N-M_E)^2$; the sum runs over all the new doublets and
we have neglected the effects of the light--heavy neutrino mixing.
The value of the $\rho$ parameter is constrained by the electroweak data.
For $m_t=174\pm16$ GeV,
as suggested by the CDF measurement \cite{cdf-mtop}, the result is
$\delta\rho=0.0004\pm0.0022\pm0.002$ \cite{rhoparam}
(the second error is from the uncertainty in the Higgs mass $m_H$)
for the ($m_t$-independent) corrections due to possible new physics.
{}From Eq. (\ref{nondec10}) and for $m_H<1$ TeV, we then find
that any new lepton doublet should be degenerate within
$\vert M_N-M_E\vert \lessim220$ GeV at 90\% C.L..
This constraint is significant if the new neutrinos are close to the
perturbative limit $M_N\lessim1$ TeV, that holds for $N$ belonging to an SU(2)
doublet, but in any case within this region it
does not require an important fine-tuning.

A second phenomenological constraint on heavy ordinary neutrinos, belonging
to weak doublets, comes from the limit on the $S$ Peskin--Takeuchi
\cite{heavyloops} parameter.
The contribution from a multiplet of heavy degenerate fermions is
$\Delta S = C{\sum_f}(t_{3L}(f)-t_{3R}(f))^2/3\pi$, where $t_{3L,R}(f)$ is
the weak isospin of the left- (right-) handed component of fermion $f$, and
$C$ is the number of colours \cite{langacker-erler}.
Then the contribution from the set of all the particles in each new ordinary
family of heavy fermions is $\Delta S\simeq 2/3\pi>0$.
{}From the analysis of the electroweak data, one gets the 95\% C.L. bound
$S<0.2$, which can be relaxed to $S<0.4$ if only
positive contributions to $S$ \cite{langacker-erler} are allowed.
As a consequence, only a single, or very marginally two, new
families are allowed to exist by the present data.
This is the maximum number of pairs $N$, $N'$, for $N$
belonging to a new family and $N'$ isosinglet, and in this case $M_{\nu N'}$
and $M_{NN'}$ can be $3\times1$ (much less likely $3\times2$) matrices.
However, in the following we will retain the general notation, allowing for
an arbitrary number of $N$, $N'$ pairs. In fact, this number is not
important for our discussion (provided it is non-zero), and
in addition one can also
consider new pairs $N$, $N'$ from (respectively) an isodoublet and
an isosinglet which do not belong to new ordinary families. An example of
the latter situation is given by E$_6$ models themselves
\cite{unconventional} that contain new leptonic doublets and isosinglets
in the ${\bf 27}$ representation, so that in the
three-family models there are also three such pairs. In this case, the
contribution to the $S$ parameter is zero, since the non-isosinglet new
states appear in vector doublets ($t_{3L}(f)=t_{3R}(f)$).

In the case when both $N$ and $N'$
are isosinglets, as in the model \cite{mohapatra-valle},
the $S$ parameter is not affected, while the contribution to $\delta\rho$
is suppressed by the fourth power in the mixing angles and is
negligible \cite{cheng-li,nuloopfcons}.

\section{Constraints from Flavour-Changing (FC) processes}

The indirect limits presented in Table 1
are more stringent than any direct bound on the
tree-level effects of the off-diagonal mixings, such as the constraints
from the search for neutrino oscillations \cite{ll2}.
However, loop diagrams involving heavy neutrinos give rise to unobserved rare
processes such as $\mu\to e \gamma$, $\mu\to ee^+e^-$, $\tau\to
l_al_b^+l_c^-$ ($a,b,c=e,\mu$), $Z\to l_a^-l_b^+$ ($a,b=e,\mu,\tau$), etc.
\cite{lim-inami,concha-valle}.
Taking into account the stringent constraints in Table 1,
the rates for all the processes involving the violation of the $\tau$
lepton number turn out to be below the experimental sensitivity even for
extreme values of the heavy neutrino masses \cite{concha-valle}.
In other words, it is not possible to improve the limits in Table 1
on the parameters $(GG^\dagger)_{a\tau}$ ($a=e,\mu$).
However the extraordinary sensitivity of the experiments looking for
FC processes involving the first two families implies that the constraints
from $\mu\to e\gamma$, from $\mu\to ee^+e^-$, or
from $\mu$--$e$  conversion in nuclei,
are significant and turn out to be stronger than those from Table 1.
The diagrams contributing to these processes are proportional to factors
involving the light--heavy mixing angles, and in the cases of the processes
$\mu\to ee^+e^-$ and $\mu$--$e$  conversion in nuclei, they
depend up to quadratically in the heavy neutrino mass scale $M$.
In see-saw models, where the mixing angles
are suppressed by inverse powers of the heavy masses, the resulting
dependence is $\sim M^{-2}$ \cite{lim-inami,cheng-li},
in agreement with the decoupling theorem \cite{decoupling}.
However, the models discussed in section 2 predict a finite light--heavy
mixing independent of the light-to-heavy mass ratio, so that,
assuming that the $N'$ states are isosinglets, a genuine
$\sim M^2$ dependence is obtained. This non-decoupling behaviour is
comparable to the top mass dependence of the $\rho$ parameter
\cite{rho-mt} and of the $Z\to b\bar b$ vertex \cite{zbb-mt}.
Since the effects of any SU(2)-invariant mass term should decouple when the
mass term goes to infinity \cite{decoupling},
in all these cases the relevant combinations
of the mass and mixing parameters entering the graphs are connected to the
electroweak breaking scale and cannot exceed $\sim1$ TeV.
This consideration is obvious for the top-dependent non-decoupling effects,
and will be explicitly verified in the following in the
case of the heavy-neutrino contributions.

\bigskip

\noindent {\it 4.1 $\mu\to e\gamma$ }

Let us first consider
the decay $\mu\to e \gamma$, induced by one-loop graphs involving virtual
heavy neutrinos \cite{ma-pramudita,ll2}.
The corresponding branching ratio is given by
\beq
B(\mu\to e \gamma)={3\alpha\over 8\pi}\left|\sum_i G_{ei}G^\dagger_{i\mu}
\phi\left(M_i^2\over M_W^2\right)\right|^2
,\label{nondec11}
\eeq
where $M_i$ is the mass of the heavy eigenstate $\N_{i}$, and
the function
\beq
\phi(x)={x(1-6x+3x^2+2x^3-6x^2\ln x)\over 2(1-x)^4}
\label{nondec12}
\eeq
varies slowly from $0$ to $1$ as $x$ ranges from $0$ to $\infty$.
Taking into account the 90\% C.L. limit $B(\mu\to e \gamma)<4.9\times10^{-
11}$ \cite{muegamma-exp}, and assuming $M_i\gtrsim M_W$, one gets the
estimate
$\vert (GG^\dagger)_{e\mu}\vert \lessim 0.95\times10^{-3}$ \cite{ll2}. This
bound holds under the assumption that no important fine-tuning works in the
sum $\sum_i G_{ei}G^\dagger_{i\mu}\phi\left(M_i^2\over M_W^2\right)$, and is
independent of the weak isospin of the new states.
We will be interested in the case of very
heavy neutrinos, $M_i\gg M_W$. In this case $\phi\simeq 1$ and we get the
stringent bound
\beq
\vert (GG^\dagger)_{e\mu}\vert \lessim 0.24\times 10^{-3}.
\label{nondec13}
\eeq
In the models discussed in Section 2,
allowing for large light--heavy mixings not suppressed by see-saw relations,
the loop contribution of Eq. (\ref{nondec11})
does not vanish for large heavy-neutrino
masses; however, it does not have a hard non-decoupling
dependence on the heavy mass scale, as
the $Ze\mu$ vertex and the box diagrams that we will discuss in the
next paragraphs.

\bigskip

\noindent {\it 4.2 $Ze\mu$ vertex }

Let us consider now the FC $Z\bar e\mu$ current,
parameterized in the form
\beq
J^\mu_{Z\bar e\mu}= g \bar e \gamma^\mu(k_LP_L+k_RP_R)\mu=
{g\over2}\bar e \gamma^\mu(k_V-k_A\gamma^5)\mu,
\label{nondec14}
\eeq
where $g=(4\sqrt2 G_F M_Z^2)^{1/2}$ is the weak coupling constant
to the $Z$ boson, and $P_{R,L}=(1\pm\gamma_5)/2$.
The leading contribution from heavy neutrinos $\N_i$ of mass $M_{i}\gg M_W$
arises from the (convergent) loop diagram in Fig. 1 involving the
exchange of {\it longitudinal} $W$ bosons (more precisely of the would-be
Goldstone bosons in the Feynman gauge).
Notice that, as far as we are
interested only on the quadratic term in the mass of the heavy states,
it is consistent to ignore all the other loop contributions to the vertex,
since their leading quadratic terms sum to zero when $M_i\gg M_W$.
We also remark that the corrections to our
approximation, which would become the main contribution for $M_i\lessim100$
GeV,
would give a phenomenologically small result due to the
constraint coming from $\mu\to e\gamma$, Eq. (\ref{nondec11}), which
would be dominant for such relatively light new states.
A more formal justification for considering only the graphs in Fig. 1 can
be given in the effective lagrangian approach \cite{efflagr}.

Neglecting the external momenta, and for $M_i\gg M_W$, the contribution in
Fig. 1 reads
\beq
k_L=k_V=k_A=-{ g^2 \over 128 \pi^2 M_Z^2}
{\cal F}_{e\mu},
\label{nondec15}
\eeq
where the dependence from the new physics parameters is given by the factor
\beq
{\cal F}_{e\mu}\equiv \sum_{i,j=heavy} S_{ij} M_i M_j f(M_i,M_j), \qquad
f(M_i,M_j) = {M_i M_j \ln (M_i^2/M_j^2) \over M_i^2-M_j^2}.
\label{nondec16}
\eeq
The dependence on the light--heavy mixing angles is contained in the term
\beq
S_{ij}\equiv G^*_{\mu i}G_{ej}
[(G^\dagger G)_{ji} + 2 t_3^{N} (H^\dagger H)_{ji}]
= G^*_{\mu i}G_{ej}
[(G^\dagger G)_{ji}(1 - 2 t_3^{N}) +  2 t_3^{N}\delta_{ji}]
,\label{nondec17}
\eeq
where $t_3^{N}$ is the weak isospin of the (left-handed) $N$ field.
The dependence is quadratic in the light--heavy mixing matrix $G$, unless
the new neutrinos $N$ are weak isosinglets, in which case it is quartic.

The best limit on the FC current $J^\mu_{Z\bar e\mu}$ arises from the search
for $\mu$--$e$  conversion in nuclei \cite{mue-exp,mue-new}.
For general FC couplings $k_V$ and $k_A$
and for nuclei with atomic number $A\lessim100$,
the induced branching ratio with the total nuclear muon capture
rate is \cite{mueconv}
\beq
R\simeq
{G_F^2\alpha^3\over\pi^2} m_\mu^3p_eE_e {Z_{eff}^4\over Z}
\vert F(q)\vert^2{1\over \Gamma_{capture}}
(k_V^2+k_A^2)Q_W^2 ,
\label{nondec18}
\eeq
where $p_e$ ($E_e$) is the electron momentum (energy), $E_e\simeq p_e\simeq
m_\mu$ for this process, and $F(q)$ is the nuclear form factor,
as measured for example from electron scattering \cite{escatt}.
Here $Q_W = (2Z+N)v_u + (Z+2N)v_d$
is the coherent nuclear charge
associated with the vector current of the nucleon, as a function of the
quark couplings to the $Z$ boson and of the nucleon charge and atomic
$(A=Z+N)$ numbers, and $Z_{eff}$ has been determined in the
literature \cite{zeff}.
For $\Gamma_{capture}$ in $^{48}_{22}$Ti we will use the
experimental determinations
$\Gamma_{capture}\simeq (2.590\pm0.012)\times10^6 {\rm s}^{-1}$
\cite{mue-exp}, $F(q^2\simeq -m_\mu^2)\simeq0.54$
and $Z_{eff}\simeq17.6$.
The resulting limit for the FC couplings is then \cite{mueconv}
\beq
(k_V^2+k_A^2) < 5.2\times10^{-13} \left(B\over 4\times10^{-12}\right),
\label{nondec19}
\eeq
where $B$ is the value of the experimental bound to $R$, $B=4\times10^{-12}$
at present \cite{mue-exp}.

Comparing with the prediction of our model, Eqs.
(\ref{nondec15})--(\ref{nondec17}),
we find that the non observation of the process of $\mu$--$e$  conversion in
nuclei results in the bound
\beq
{\vert {\cal F}_{e\mu}\vert \over({\rm 100 GeV})^2}\lessim0.97\times10^{-3}
\left(B\over 4\times10^{-12}\right)^{1/2}.
\label{nondec20}
\eeq

In order to find out the impact of this constraint, we have to specify
the value of the weak isospin of the new states involved in the mixing.

Let us consider first the case i) of Section 2,
when the new states $N$ are
{\it ordinary}, that is $t_3^{N}=1/2$. In this case, a
substantial light--heavy mixing can be expected, since
$M_{\nu N'}$ and $M_{N N'}$ can be generated both by the VEVs of
SU(2)-doublet Higgs fields (we are assuming that $N'$ are singlets).
Then $S_{ij}=G_{\mu i}^*G_{e i} \delta_{ij}$, and only the diagonal terms
contribute in Eq. (\ref{nondec17}).
It is easy to show that $f(M,M)=1$, then Eq. (\ref{nondec16})
simplifies to
\beq
{\cal F}_{e\mu}=(G M^2 G^\dagger)_{e\mu}=
(M_{\nu N'}M^\dagger_{\nu N'})_{e\mu},
\label{nondec21}
\eeq
where $M$ is the diagonal Dirac mass matrix for the heavy states appearing
in Eq. (\ref{nondec3}), and we have used Eqs.
(\ref{nondec2})--(\ref{nondec4}).
Since $M_{\nu N'}$ and $M_{N N'}$ arise from the
breaking of the SU(2) symmetry, their entries are expected to be
generated at the electroweak scale.
In particular, the heavy masses should be $M_i\lessim1$ TeV, assuming
perturbation theory not to be spoiled. We see explicitly in this case that
the non-decoupling behaviour of the $Ze\mu$ vertex is due to the breaking
of the SU(2) symmetry.
On the other hand, since
${\cal F}_{e\mu}=(M_{\nu N'}M^\dagger_{\nu N'})_{e\mu}$ is naturally
expected to be $\sim (100$ GeV$)^2$, we see that Eq. (\ref{nondec20})
indeed represents a strong constraint on the model, like the bounds
on the mixing matrix $GG^\dagger$ discussed in the previous section and the
limit from $\mu\to e \gamma$ of the previous paragraph.

To compare these different bounds, let us assume for
simplicity that the mass differences
amongst the heavy states are smaller than their common scale $M$, namely
$\vert M_i^2-M_j^2\vert/(M_i^2+M_j^2)\ll1$.
The allowed regions in the LFV mixing
parameter $S\equiv (GG^\dagger)_{e\mu}^{1/2}$
and the heavy mass scale $M$ is then
given in Fig. 2. For $200$ GeV$\lessim M$,
the constraint on $S$ from $\mu$--$e$  conversion in nuclei,
given by Eq. (\ref{nondec20}) and represented by the full line in Fig.
2\footnote{For $M\to 100$ GeV, towards the left margin of the figure,
the non-decoupling contribution
to the $Ze\mu$ vertex is not more important than the others we have
neglected \cite{efflagr},
but in this regime the bound from $\mu\to e \gamma$ dominates
\cite{lim-inami,concha-valle}.},
is more stringent than the bound from $\mu\to e\gamma$ (dashed line),
resulting from Eqs. (\ref{nondec11}), (\ref{nondec12}) and Ref.
\cite{muegamma-exp}.
If the new states are lighter than $\sim200$ GeV,
the constraint from $\mu\to e \gamma$ is the most stringent one.
The indirect limit from Table 1, $S=(GG^\dagger)_{e\mu}^{1/2}
<\sqrt{0.004}=0.063$,
is much worse and would be represented by a horizontal line above the
figure.

Let us now consider the case ii) of Section 2,
in which the new states $N$ mixing with the
light neutrinos are singlets under SU(2).
Then from Eq. (\ref{nondec17})
we see that the mixing factor depends of the fourth power of the
light--heavy mixings. One could expect that in this case no significant
constraint can be obtained from Eq. (\ref{nondec20});
however the mass eigenvalues $M_i$ are no longer limited by 1 TeV.
In fact, since the states $N'$ are also assumed to be isosinglets,
the entries $M_{NN'}$ in Eq. (\ref{nondec2}) are not related to the
electroweak scale and can be expected to be generated at a higher scale.
To have an idea of the impact of the constraint of Eq.
(\ref{nondec20}) in this case, let us assume again
that the mass differences
amongst the heavy states are smaller than their common scale $M$.
In this case,
$f(M_i,M_j)\simeq1$, and using the identity $MG^\dagger=K^TM^T_{\nu N'}$
(which can be deduced from Eqs. (\ref{nondec3}) and (\ref{nondec4})),
we find that
${\cal F}_{e\mu}\simeq (GMG^\dagger)^2_{e\mu}=
(G_{ej}G^\dagger_{i\mu })(K^\dagger M_{\nu N'}^\dagger
M_{\nu N'}K)_{ij}$.
Since $K$ is unitary, we can expect that
$\vert (K^\dagger M_{\nu N'}^\dagger M_{\nu N'}K)_{ij}\vert\sim \bar
M_{\nu N'}^2$, where $\bar M_{\nu N'}$ is an average scale for the
entries of the SU(2)-breaking
mass matrix $M_{\nu N'}$, which is generated at the
electroweak scale. Again, we see explicitly that the non-decoupling
behaviour is due to the spontaneous breaking of the SU(2) symmetry.

A more drastic approximation,
${\cal F}_{e\mu}\sim(GG^\dagger)_{e\mu} \bar M_{\nu N'}^2$,
gives an expression that is
similar to Eq. (\ref{nondec21}), corresponding to the mixing with
doublet neutrinos. In fact,
$\bar M_{\nu N'}\lessim1$ TeV since it breaks SU(2),
so that the considerations given in the
previous case of doublet new neutrinos can be repeated here.
The constraint
from $\mu$--$e$  conversion in nuclei can be represented again by
the full line in Fig. 2, after the substitution $M\to \bar M_{\nu N'}$.
In particular, the constraint of Eq. (\ref{nondec20})
is more stringent than the constraint from $\mu\to e\gamma$
if $\bar M_{\nu N'}\gtrsim200$ GeV.

If $\bar M_{\nu N'}$ and $\bar M_{N' N'}\simeq M$ are the typical orders of
magnitude of the entries of the matrices $ M_{\nu N'}$ and
$M_{N' N'}$, from Eq. (\ref{nondec4b})
the typical value of the mixing angles is
\beq
S\equiv(GG^\dagger)_{e\mu}^{1/2} \sim \bar M_{\nu N'}/M,
\label{nondec23}
\eeq
which is of course similar to the see-saw formula
\cite{see-saw,cheng-li}, though no longer related to the ratio of the
physical mass eigenstates.
Then the range $200$ GeV $\lessim \bar M_{\nu N'}\lessim 1$ TeV
and the bound of Eq. (\ref{nondec20}) correspond to a heavy scale $M
\gtrsim(10-300)$ TeV. This last constraint is not problematic, since
in the present case we are considering singlet new states which can be
originated by SU(2)-invariant VEVs. In fact, the model {\it predicts}
observable LFV effects if the latter inequality is
(almost) an equality, corresponding to the
existence of an intermediate scale $\lessim 300$ TeV for the heavy states.
However, for the mixing with singlet neutrinos this assumption on
the heavy scale would be somewhat arbitrary (unless a justification is
given by fully specifying the model), so that in general
significant LFV effects are not necessarily
{\it predicted} in this case, contrary to the case of the mixing with
new {\it ordinary} states that we have considered above.

Moreover, Eq. (\ref{nondec23}) allows us to express the dependence of
the leading contribution to the $\mu$--$e$  vertex in terms
of the mass parameters alone. In fact, for the mixing
with singlet neutrinos we get
${\cal F}_{e\mu}\sim  (\bar M_{\nu N'}^2/M)^2$,
which becomes vanishingly small when the invariant mass term
$M\to \infty$, since $\bar M_{\nu N'}\lessim 1$ TeV.
This can be considered as a
generalization of the decoupling theorem \cite{decoupling,cheng-li}
to the case of the mixing with singlet neutrinos $N$ in
the class of models characterized by Eq. (\ref{nondec2}).

\bigskip

\noindent {\it 4.3 $\mu\to ee^+e^-$ }

The leading contribution from heavy neutrinos $\N_i$ of mass $M_{i}\gg M_W$
arises from the (convergent) loop diagrams in Fig. 3 involving the
exchange of {\it longitudinal} $W$ bosons (more precisely of the would-be
Goldstone bosons in the Feynman gauge). Again, since we are
interested only in the coefficient of the quadratic term in the masses $M_i$
of the heavy states, it is consistent to ignore all the other loop
contributions to the process \cite{lim-inami,cheng-li,concha-valle}.
Neglecting the external momenta, and for $M_i\gg M_W$, we obtain for the
branching ratio, compared to the main channel $\mu\to e\nu\bar \nu$:
\beq
{B(\mu\to e e^-e^+)\over B(\mu\to e \nu\bar\nu)}=
8 \left( g^2 \over 16^2 \pi^2 M_Z^2 \right)^2
\left(\vert {\cal B}_{e\mu} - 2 \epsilon_L {\cal F}_{e\mu} \vert^2
+ {1\over2} \vert 2 \epsilon_R {\cal F}_{e\mu} \vert^2 \right),
\label{nondec24}
\eeq
where $\epsilon_L=-1/2+s_w^2\simeq-0.27$ and
$\epsilon_R=s_w^2\simeq0.23$ are the
SM left- and right- handed current couplings of the electron to the $Z$.
The mixing factor entering the box diagram is
\beq
{\cal B}_{e\mu}\equiv
\sum_{i,j=heavy} G^*_{\mu i}G_{ei} G^*_{e j}G_{ej}
M_i M_j f(M_i,M_j),
\label{nondec25}
\eeq
while ${\cal F}_{e\mu}$ and the loop integral $f(M_i,M_j)$,
entering also in the $Ze\mu$ vertex, are given by Eqs. (\ref{nondec16})
and (\ref{nondec17}).

The 90\% C.L. experimental bound, $B(\mu\to e e^-e^+)<1.0 \times 10^{-12}$
\cite{mueee-exp}, then results in the limit
\beq
{\left(\vert {\cal B}_{e\mu} + 0.54 {\cal F}_{e\mu} \vert^2
+ {1\over2} \vert 0.46 {\cal F}_{e\mu} \vert^2 \right)^{1/2}
\over (100 \, {\rm GeV})^2}
< 1.4\times 10^{-3} \left( B\over 10^{-12}\right)^{1/2}
.\label{nondec26}
\eeq
This constraint is complementary to the limits from $\mu\to e\gamma$ and
from $\mu$--$e$  conversion in nuclei, Eqs.
(\ref{nondec13}) and (\ref{nondec20}),
since it affects a different combination of the mixing parameters, namely
${\cal B}_{e\mu}$.
As a general result, we see that the contribution to
the amplitude for $\mu\to e e^+e^-$ presents a
leading quadratic dependence on the heavy mass scale, similar to that
of the $Ze\mu$ vertex.

In the case i) of section 2, when the mixing is with
new ordinary neutrinos, the vertex contribution of Fig. 3.a depends
quadratically on the light--heavy mixing angles, so that we can neglect the
the box diagram 3.b, which depends on the fourth power in the mixings.
Then the constraint of Eq. (\ref{nondec26}) becomes
\beq
{\vert {\cal F}_{e\mu}\vert \over({\rm 100\, GeV})^2}
\lessim2.3\times 10^{-3}
\left(B\over 10^{-12}\right)^{1/2}.
\label{nondec28}
\eeq
We find that in this case the limit of Eq. (\ref{nondec20})
from $\mu$--$e$  conversion in nuclei is stronger by a factor
$\sim2$, as could be expected from Ref. \cite{mueconv}.

For this reason, the limit from $\mu\to e e^+e^-$ can be important
only in the case ii) of section 2,
when the new states $N$ involved in the mixing are
singlets, since in the opposite case the contribution to the $Ze\mu$ vertex
is quadratic in the mixing angles and the bound from $\mu$--$e$
conversion in nuclei results in a stronger
constraint. When the mixing is with new singlets $N$,
the two general constraints, Eqs. (\ref{nondec20}) and (\ref{nondec26}),
appear to
be of similar strength. For a more quantitative confrontation, let us
consider again a particular case, when the mass differences
amongst the heavy states are smaller than their common scale,
so that $f(M_i,M_j)\simeq1$. In this case, we have
${\cal B}_{e\mu}\simeq (GMG^\dagger)_{ee}(GMG^\dagger)_{e\mu}$, while
${\cal F}_{e\mu}\simeq (GMG^\dagger)^2_{e\mu}=
(GMG^\dagger)_{ee}(GMG^\dagger)_{e\mu}+
(GMG^\dagger)_{e\mu}(GMG^\dagger)_{\mu\mu}+
(GMG^\dagger)_{e\tau}(GMG^\dagger)_{\tau\mu}$.
If $(GMG^\dagger)_{ee}(GMG^\dagger)_{e\mu}$
is the largest contribution to ${\cal F}_{e\mu}$, then
${\cal F}_{e\mu}\simeq{\cal B}_{e\mu}\simeq
(GMG^\dagger)_{ee}(GMG^\dagger)_{e\mu}$, and the constraint of Eq.
(\ref{nondec26}) becomes
\beq
{\vert {\cal F}_{e\mu}\vert \over({\rm 100\, GeV})^2}
\lessim0.93\times 10^{-3}
\left(B\over 10^{-12}\right)^{1/2},
\label{nondec27}
\eeq
which is as stringent as the bound from $\mu$--$e$  conversion in nuclei,
Eq. (\ref{nondec20}).
On the other hand, if $(GMG^\dagger)_{ee}(GMG^\dagger)_{e\mu}$ is
not the main part of ${\cal F}_{e\mu}$, e.g.
$(GMG^\dagger)_{ee}(GMG^\dagger)_{e\mu}<
(GMG^\dagger)_{e\tau}(GMG^\dagger)_{\tau\mu}$, then
${\cal F}_{e\mu}\gtrsim{\cal B}_{e\mu}$. If ${\cal B}_{e\mu}$ can be
neglected, the rate for $\mu\to ee^+e^-$ is given
mainly by the $Ze\mu$ vertex contribution, and
the constraint of Eq. (\ref{nondec26}) is given again by
Eq. (\ref{nondec28}) and is less important by a factor $\sim2$ than
the limit of Eq. (\ref{nondec20}) from $\mu$--$e$  conversion in nuclei.

\section{Conclusions}

We have considered a class of models predicting new heavy
neutral fermionic states, whose mixing with the light neutrinos can
be naturally significant. In contrast with the see-saw models,
the known neutrino masses are predicted to vanish due to
a symmetry, such as lepton number.
Possible non-vanishing masses for the light neutrinos could then be
attributed to small violations of such a symmetry.
We have then reviewed the bounds on the flavour-diagonal
light--heavy mixing parameters, arising mainly from the constraints on Charged
Current Universality and the LEP data, and we have updated and
collected in Table 1
the indirect limits on the flavour non-diagonal mixing parameters.

In spite of these stringent constraints,
the one-loop-induced rare processes due to the exchange of virtual
heavy neutrinos, involving the violation of the muon and electron lepton
numbers, which is tested with an impressive experimental precision,
have then been shown to be potentially significant.
In particular, the $Ze\mu$ vertex, constrained by the non-observation of
$\mu$--$e$ conversion in nuclei, and the amplitude for $\mu\to ee^+e^-$, show a
non-decoupling quadratic dependence on the heavy neutrino mass $M$, while
$\mu\to e\gamma$ is almost independent of the heavy mass above the
electroweak scale. These three processes are then used to set constraints on
the LFV parameters entering the corresponding loop diagrams, which turn
out to be stronger than the indirect limits of Table 1.

If the mass scale $M$ of the heavy states is in the range
$M_Z\lessim M\lessim 200$ GeV, the best constraint on the light--heavy LFV
mixing comes from $\mu\to e\gamma$. If $M\gtrsim200$ GeV, and the heavy
neutrinos involved in the mixing are not singlets under SU(2), the
best constraint is given in most cases
by the present data on the search for $\mu$--$e$  conversion in nuclei, and the
planned experiments looking for this process have an opportunity to find
signals from this kind of physics. In fact, we have pointed out that
in this case a significant rate
is naturally {\it expected} in the class of models considered here.

In contrast, if the heavy neutrinos mixing with the light states are
singlets under SU(2), then the leading contribution to
the $Ze\mu$ vertex is suppressed by two more powers of the light--heavy
mixing angles, and the constraint from $\mu$--$e$  conversion in nuclei is
comparable to that from $\mu\to ee^+e^-$. These two bounds turn out to be
important in a region of the parameter space corresponding to
SU(2)-breaking mass entries in the range $200\, {\rm GeV}\lessim
M_{\nu N'}\lessim1\, {\rm TeV}$.
In contrast to the model with non-singlet heavy states involved in the
mixing, in this latter case
of the mixing with isosinglet neutrinos no general prediction can be made
on the heavy mass scale, and our constraints are significant only
if it lies at an `intermediate' scale $M\lessim300$ TeV.
Moreover, in this particular case the prediction for the LFV
observables decreases for increasing values
of the heavy mass scale $M\to\infty$,
resulting in a generalization of the decoupling theorem
to this class of `non-see-saw' models.

In all the cases considered, we have explicitly discussed how the
non-decoupling behaviour is strictly related to the spontaneous breaking
of the SU(2) symmetry. This result could be expected, since
the Appelquist--Carazzone theorem \cite{decoupling}
applies to the unbroken gauge theory.

\vskip 1.truecm

\begin{center}
{\bf ACKNOWLEDGMENTS}
\end{center}

We are grateful to A. Santamaria, M.C. Gonzalez-Garcia,
E. Nardi, S. Peris, N. Rius, E. Roulet and J. Peltoniemi for several
very useful discussions and for critically reading the
preliminary versions of the paper.

\newpage

\begin{table}[p]
\begin{center}
\begin{tabular}{|c||c|c|c|}
\hline
 & $e$ & $\mu$& $\tau$ \\
\hline
\hline
$e$   & 0.007(0.005) & 0.004(0.003) & 0.015(0.004) \\
\hline
$\mu$ & 0.004(0.003) & 0.002        & 0.007(0.004) \\
\hline
$\tau$& 0.015(0.004) & 0.007(0.004) & 0.03(0.01)   \\
\hline
\end{tabular}
\vskip 2cm
\caption{The 90\% C.L. upper bound on the entries of
the matrix $GG^\dagger$ describing the effects of the
light--heavy neutrino mixing. The stronger limits in the parentheses
correspond to the mixing with isosinglet neutrinos.}
\end{center}
\end{table}

\begin{figure}
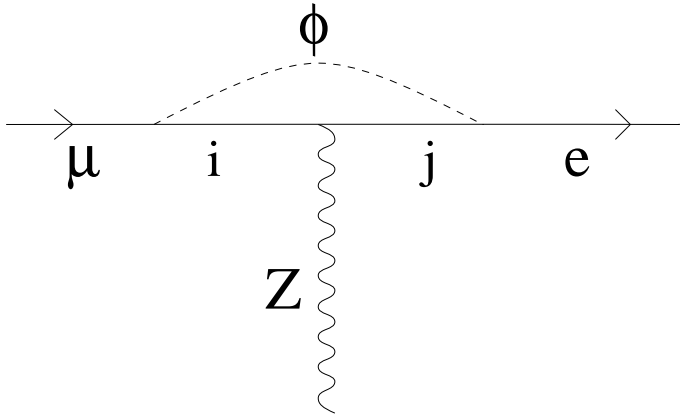

\epsfxsize = 9cm
\leavevmode
\vskip3cm
\caption{One-loop diagram for the $Ze\mu$ vertex due to virtual heavy
neutrinos $\N_{i,j}$ and would-be Goldstone boson $\phi^\pm$,
representing the Landau gauge leading contribution in the limit
$M_{i,j}\gg M_Z$. }
\end{figure}

\begin{figure}
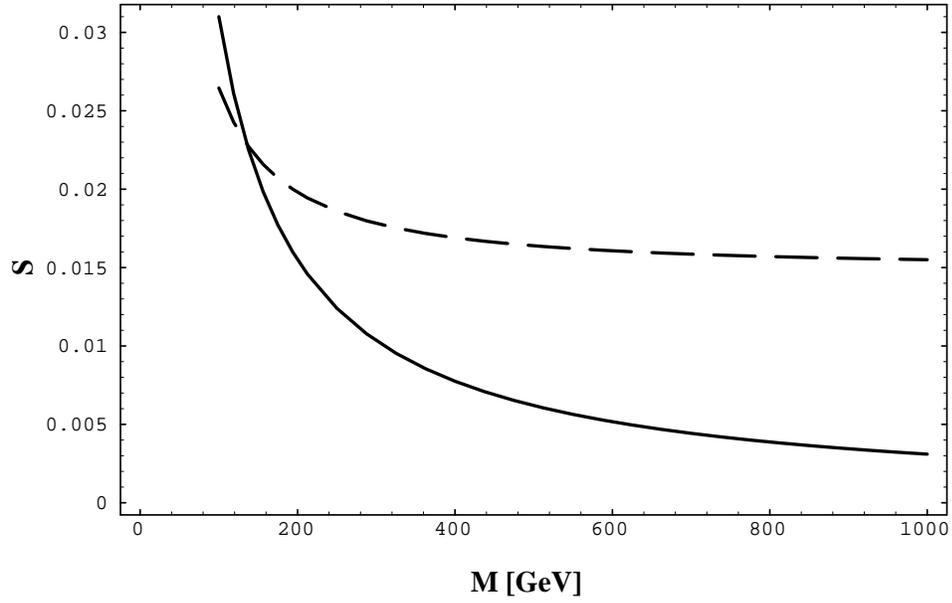

\begin{center}
\epsfxsize = 13cm
\leavevmode
\end{center}
\caption{90 \% C.L. allowed regions for the LFV mixing $S\equiv
(GG^\dagger)_{e\mu}^{1/2}$ versus the heavy neutrino mass scale $M$,
from the limits on $\mu\to e\gamma$ (dashed line) and on
$\mu$--e conversion in nuclei (full line). }
\end{figure}

\begin{figure}
\begin{center}
\epsfxsize = 13cm
\leavevmode
\end{center}
\vskip4cm
\caption{One-loop diagrams for $\mu\to e e^+e^-$ due to virtual heavy
neutrinos $\N_{i,j}$ and would-be Goldstone boson $\phi^\pm$,
representing the Landau gauge leading contributions in the limit
$M_{i,j}\gg M_Z$. }
\end{figure}


\newpage

\begin{thebibliography}{99}
\frenchspacing

\bibitem{fermionmix}
P. Langacker and D. London, \pr{D38} (1988) 886; \\
E. Nardi and E. Roulet, \pl{B248} (1990) 139; \\
G. Bhattacharyya \ea, \prl{64} (1990) 2870; \\
G. Bhattacharyya \ea, \mpl{A6} (1991) 2921; \\
E. Nardi, E. Roulet and D. Tommasini, \np{B386} (1992) 239; \\
E. Nardi, E. Roulet and D. Tommasini, \pr{D46} (1992) 3040; \\
C. Cs\'aki and F. Csikor, \pl{B309} (1993) 103;\\
C.P. Burgess \ea,  \pr{D49} (1994) 6115; \\
G. Bhattacharyya, \pl{B331} (1994) 143.

\bibitem{fitnu} E. Nardi, E. Roulet and D. Tommasini,
\pl{B327} (1994) 319.

\bibitem{mix94} E. Nardi, E. Roulet and D. Tommasini,
\pl{B344} (1995) 225.

\bibitem{see-saw}
T. Yanagida, \ptp{B135} (1978) 66;\\
M. Gell-Mann, P. Ramond and R. Slansky, in {\it Supergravity}, eds.
P. Van Nieuwenhuizen and D. Freedman (North-Holland, Amsterdam, 1979), p. 315.

\bibitem{lim-inami}
J. Bjorken, K. Lane and S. Weinberg, \pr{D16} (1977) 1474;\\
T.P. Cheng and L.F. Li, \pr{D16} (1977) 1425;\\
C.S. Lim and T. Inami, \ptp{67} (1982) 69;\\
J.D. Vergados, \prep{133} (1986) 1.

\bibitem{cheng-li}
T.P. Cheng and L.F. Li, \pr{D44} (1991) 1502.

\bibitem{numix}
D. Wyler and L. Wolfenstein, \np{B218} (1983) 205.

\bibitem{mohapatra-valle}
R.N. Mohapatra and J.W.F. Valle, \pr{D34} (1986) 1642; \\
E. Witten, \np{B268} (1986) 79;\\
J. Bernab\'eu et al., \pl{B187} (1987) 303; \\
J.L. Hewett and T.G. Rizzo, \prep{183} (1989) 193.

\bibitem{ll2}
P. Langacker and D. London, \pr{D38} (1988) 907.

\bibitem{unconventional}
E. Nardi, \pr{D48} (1993) 3277.

\bibitem{pdg94}
Rev. of Part. Properties, \pr{D50} (1994) 1173.

\bibitem{buchmuller-greub}
W. Buchm\"uller and C. Greub, \np{B363} (1991) 345.

\bibitem{maalampi-roos}
J. Maalampi and M. Roos, \prep{186} (1990) 53, and references therein.

\bibitem{lep94}
The LEP Electroweak Working Group, preprint DELPHI 94-33 PHYS 364 (1994).

\bibitem{alr-slc}
SLD Collaboration, K. Abe \ea, \prl{73} (1994) 25.

\bibitem{nuloopfcons}
J. Bernab\'eu \ea, \prl{71} (1993) 2695.

\bibitem{rho-mt}
M. Veltman, \np{B123} (1977) 89.

\bibitem{rhoparam}
V. Barger \ea, \pr{D30} (1984) 947;\\
V. Barger, J.L Hewett and T.G. Rizzo, \mpl{A5} (1990) 743. \\
P. Langacker and J. Erler, in Ref. \cite{pdg94}, p. 1312.

\bibitem{cdf-mtop}
CDF Collaboration, F. Abe \ea, preprint FERMILAB-PUB-94-116-E
(1994).

\bibitem{langacker-erler}
P. Langacker and J. Erler, in Ref. \cite{pdg94}, p. 1312.

\bibitem{heavyloops}
M.E. Peskin and T. Takeuchi, \prl{65} (1990) 964; \\
W.J. Marciano and J.L. Rosner, \prl{65} (1990) 2963; \\
D.C. Kennedy  and P. Langacker \prl{65} (1990) 2967; E: ibidem {\bf 66}
(1991) 395; \\
M. Golden and L. Randall, \np{B361} (1991) 3; \\
D.C. Kennedy  and P. Langacker, \pr{D44} (1991) 1591;\\
G. Altarelli and R. Barbieri, \pl{B253} (1991) 161; \\
M.E. Peskin and T. Takeuchi, \pr{D46} (1992) 381.

\bibitem{concha-valle}
M.C. Gonzalez-Garcia and J.W.F. Valle, Mod. Phys. Lett. {\bf A7} (1992)
477; E ibidem {\bf A9} (1994) 2569;\\
A. Ilakovac and A. Pilaftsis, Rutherford preprint RAL/94-032 (1994).

\bibitem{vergados}
T.S. Kosmas, G.K. Leontaris and J.D. Vergados, Ioannina University
preprint IOA. 300/93 (1993).

\bibitem{decoupling}
T. Appelquist and J. Carazzone, \pr{D11} (1975) 2856.

\bibitem{zbb-mt}
A.A. Akhundov, D.Y. Bardin and T. Riemann, \np{B276} (1986) 1;\\
J. Bernab\'eu, A. Pich and A. Santamaria, \pl{B200} (1988) 569;\\
W. Beenakker and W. Hollik, \zp{C40} (1988) 141;\\
J. Bernab\'eu, A. Pich and A. Santamaria, \np{B363} (1991) 326.

\bibitem{ma-pramudita}
E. Ma and A. Pramudita, \pr{D24} (1981) 1410.

\bibitem{muegamma-exp}
LAMPF collaboration, R.D. Bolton \ea, \prl{56} (1986) 2461.

\bibitem{efflagr}
M. Bilenky and A. Santamaria, \np{B420} (1994) 47.

\bibitem{mue-exp}
T. Suzuki, D.F. Measday and J.P. Roalsvig, \pr{C35} (1987) 2212.

\bibitem{mue-new}
A. Badertscher \ea, SINDRUM II collaboration, Villigen preprint PSI-PR-90-41
(1990);\\
V.S. Abadjev \ea, MELC collaboration, Moscow preprint (1992).

\bibitem{mueconv}
J. Bernab\'eu, E. Nardi and D. Tommasini, \np{B409} (1993) 69.

\bibitem{escatt}
I. Sick and J.S. McCarthy, \np{A150} (1970) 631; \\
W. Bertozzi, J. Friar, J. Heisenberg and J.W. Negele, \pl{B41} (1972)
408;\\
W. Bertozzi \ea,  \prl{28} (1972) 1711;\\
B. Dreher \ea, \np{A235} (1974) 219;\\
W. Donnely and J.D. Walecka, \arns{25} (1975) 329;\\
B. Frois and C.N. Papanicolas,
\arns{37} (1987) 133.

\bibitem{zeff}
J.C. Sens, \pr{113} (1959) 679;\\
K.W. Ford and J.G. Wills, \np{35} (1962) 295;\\
R. Pla and J. Bernab\'eu, \anfis{67} (1971) 455;\\
H.C. Chiang \ea, \np{A559} (1993) 526.

\bibitem{mueee-exp}
SINDRUM collaboration, U. Bellgardt \ea, \np{B299} (1988) 1.
\end{thebibliography}
\end{document}